\documentclass[global,twocolumn,final]{svjour}
\usepackage[utf8]{inputenc}	
\usepackage[unicode]{hyperref}
\usepackage{subcaption}
\usepackage{graphicx}
\usepackage{pgf}
\usepackage[locale=US]{siunitx}
\title{External cavity diode laser setup with two interference filters}
\author{Alexander Martin \and Patrick Baus\thanks{\email{patrick.baus@physik.tu-darmstadt.de}} \and Gerhard Birkl}
\hypersetup{
	pdftitle={External cavity diode laser setup with two interference filters}, pdfauthor={Alexander Martin and Patrick Baus and Gerhard Birkl},
	colorlinks = false,
    urlbordercolor = {white}
}

\institute{Technische Universität Darmstadt, Institut für Angewandte Physik, Schlossgartenstraße 7, 64289 Darmstadt, Germany}

\journalname{Applied Physics B: Lasers and Optics}
\date{Received: 30 August 2016 / Accepted: 4 November 2016 }

\begin{document}

\maketitle

\begin{abstract}
We present an external cavity diode laser setup using two identical, commercially available interference filters operated in the blue wavelength range around \SI{450}{\nm}. The combination of the two filters decreases the transmission width, while increasing the edge steepness without a significant reduction in peak transmittance. Due to the broad spectral transmission of such interference filters compared to the internal mode spacing of blue laser diodes, an additional locking scheme, based on Hänsch--Couillaud locking to a cavity, has been added to improve the stability. The laser is stabilized to a line in the tellurium spectrum via saturation spectroscopy, and single-frequency operation for a duration of two days is demonstrated by monitoring the error signal of the lock and the piezo drive compensating the length change of the external resonator due to air pressure variations. Additionally, transmission curves of the filters and the spectra of a sample of diodes are given.
\end{abstract}

\section{Introduction}

\begin{figure}
	\centering
	\begin{subfigure}[b]{0.40\textwidth}
		\includegraphics[width=1 \linewidth]{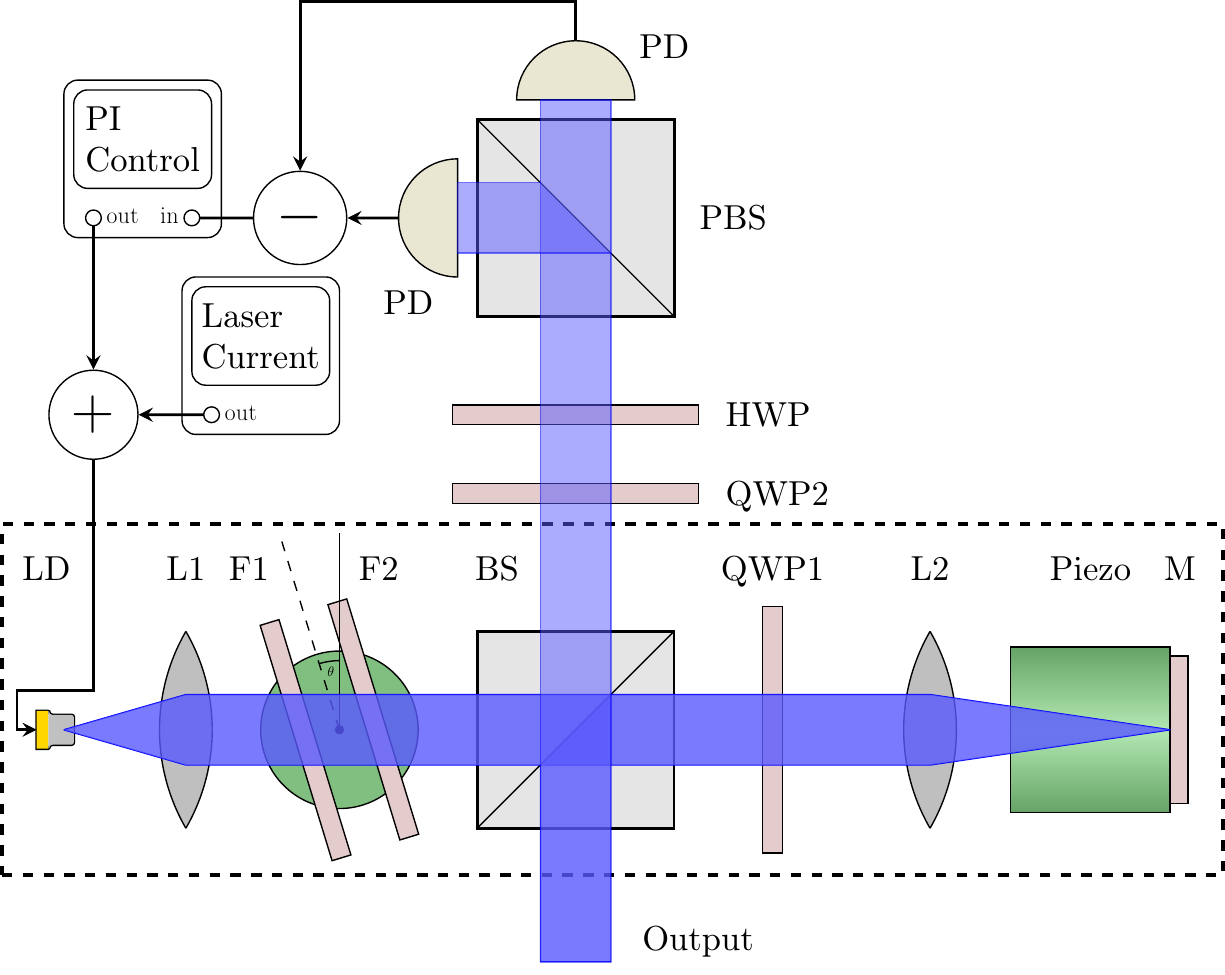}
	\end{subfigure}
	
	\begin{subfigure}[b]{0.40\textwidth}
		\includegraphics[width=1 \linewidth]{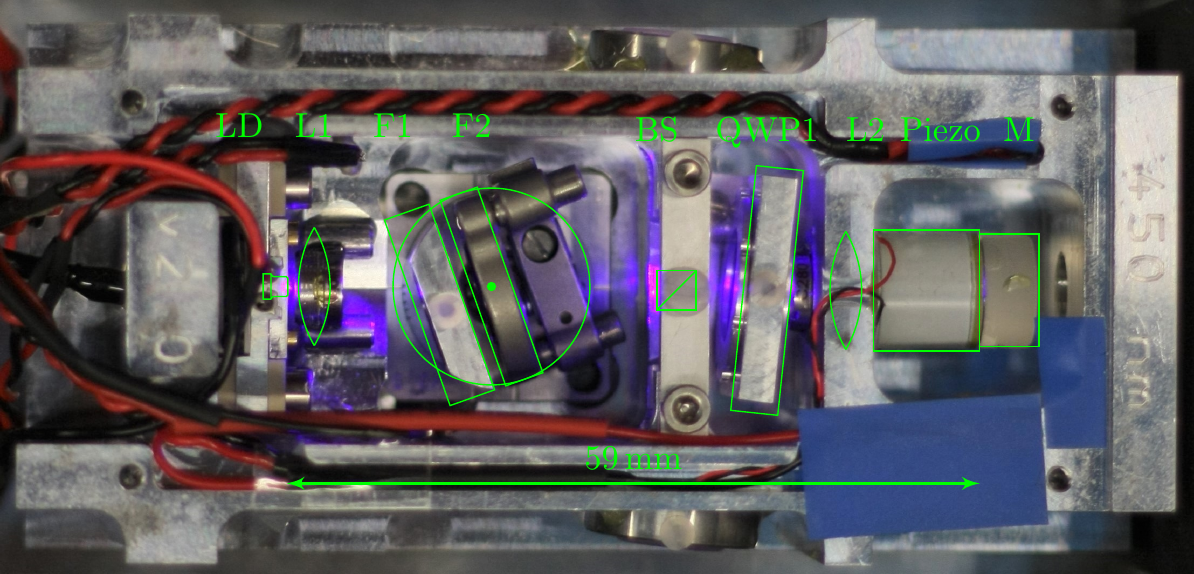}
	\end{subfigure}
	\caption{(\textit{Top}) schematic setup of the external cavity diode laser (\textit{dashed box}) with two interference filters  and the additional elements for polarization-stabilization of the cavity length. The two filters are mounted on a rotation stage for easy wavelength tuning. The external resonator has a total geometrical length of \SI{59}{\mm} from laser diode LD to mirror M. (\textit{bottom}) photograph of the laser setup}
	\label{fig:ecdl_pbs}
\end{figure}

Modern optical spectroscopy as well as the manipulation of the internal and external degrees of freedom of atoms and molecules (neutral and charged) requires stable laser sources in a wide range of wavelengths.
External cavity diode lasers (ECDLs) are an excellent choice in many cases \cite{:/content/aip/journal/rsi/62/1/10.1063/1.1142305} and are specifically used because of their low cost and the broad wavelength availability, which covers almost the entire spectrum from the near-infrared to near-ultraviolet. The first compact ECDL setup for single-frequency operation using a grating for the external cavity was described by Ted Hänsch in 1995 \cite{RICCI1995541} and has remained an important configuration ever since. Twenty years later, ECDLs with the external feedback realized by a narrow-bandwidth interference filter and a semitransparent mirror \cite{Baillard2006609} improved the stability against external perturbations.
In recent years, the development of blue laser diodes based on gallium nitride and indium gallium nitride \cite{NakamuraChichibu2000}
extended the range of ECDLs to shorter wavelength. Nevertheless, due to the specific characteristics of these laser diodes, stable single-frequency operation in the blue wavelengths region remains a challenge. 

In this work, we present an ECDL setup for stable single-frequency operation in the wavelength range around \SI{450}{\nm} (Fig.\ \ref{fig:ecdl_pbs}). We combine the design of an interference filter-based ECDL \cite{Baillard2006609} in the specific realization of Ref. \cite{Thompson2012}, adding a second filter and the intra-cavity stabilization technique presented by Führer et al.\ \cite{ThorstenFuehrer2009,Fuhrer:11}. The latter is based on the polarization-dependent locking scheme introduced by Hänsch and Couillaud \cite{Hansch1980} for locking lasers to an external cavity.
We use a PL-450B indium gallium nitride (InGaN) laser diode manufactured by OSRAM \cite{OSRAMPl450B} as active element. This single transverse mode diode has an optical output power of up to \SI{80}{\mW} in free-running mode. The high availability in volume and the low price make it an interesting candidate for ECDL developments. On the other hand, the spectral characteristics are less favorable for single-frequency operation than for more expensive diode variants as used in grating stabilized ECDLs in the blue so far \cite{ShimadaChidaOhtsuboEtAl2013,Dutta:16}. 

In the next sections, we first discuss the realization of the ECDL setup. Then, we will present the characterization of a set of free-running PL-450B laser diodes and the interference filter used. This is followed by the characterization of the complete diode laser and by the demonstration of the long-term stability of the laser locked to a molecular reference.

\section{Laser setup}
\label{sec:setup}
The laser setup depicted in Fig. \ref{fig:ecdl_pbs} is a variation of the design of Ref.\cite{Baillard2006609} in the form introduced in \cite{Thompson2012}. The light from the laser diode (LD) is collimated by a lens (L1) with \SI{3.1}{\mm} focal length. The external resonator consists of a silver coated mirror (M) (reflectivity ${\textup R_\text{M}} = \SI{99}{\percent}$) and lens L2 ($f = \SI{ 18.4}{\mm}$) in a cateye configuration. The resonator has a total length of \SI{59}{\mm}. For adjustment of the resonator length, the mirror is glued to a piezoelectric ring actuator with a length change coefficient of approximately \SI{100}{\nm/\V}.
For outcoupling of the laser light, we use a polarization-dependent beam splitter cube (BS) combined with a quarter wave plate (QWP1). The BS has a transmission of ${\textup T_\text{p}} \approx \SI{90}{\percent}$ (\SI{10}{\percent}) for p-polarized (s-polarized) light.
Depending on the rotation angle of QWP1, the back-reflectivity of the combination of BS, QWP1, and M can be varied between ${\textup T_\text{p}} \cdot {\textup R_\text{M}} \cdot {\textup T_\text{s}} \approx \SI{9}{\percent}$ and ${\textup T_\text{p}} \cdot {\textup R_\text{M}} \cdot {\textup T_\text{p}} \approx \SI{80}{\percent}$. The laser diode must therefore be mounted with its polarization axis in the p-plane of BS. 
Two band-pass filters (F1 and F2) provide the wavelength selection. We use interference filters (Semrock Laser-Line LL01-458) with a center wavelength of \SI{457.9}{\nm} and a measured full width at half maximum (FWHM) of \SI{1.6}{\nm} at normal incidence. The filters are mounted on a rotary stage for easy tuning of the transmission wavelength. If required, a nonzero relative angle ($\theta_1 \neq \theta_2$) allows the modification of the effective spectral transmission characteristics.

The central element for improved long-term stability of this laser setup is the additional intra-cavity lock based on the Hänsch--Couillaud locking scheme \cite{Hansch1980} using the fact that a mismatch between the internal modes of the laser diode gain medium and the external resonator modes changes the state of polarization of the light in the external resonator. Following the implementation of Führer et al.\ \cite{ThorstenFuehrer2009}, the second output beam produced by the beam splitter BS is used for generation of the respective error signal: The ratio between p- and s-polarization in the external resonator is measured by subtracting the photodiode signals after the extra-cavity polarizing beam splitter (PBS) as shown in Fig. \ref{fig:ecdl_pbs}. 
With QWP2 and a half-wave plate (HWP), the signal-to-noise ratio of the error signal can be maximized. A proportional-integral (PI) controller locks the internal to the external resonator modes by regulating the laser current.

\section{Experimental results}
\subsection{Laser diode and filter}

\begin{figure}
	\centering
	\includegraphics{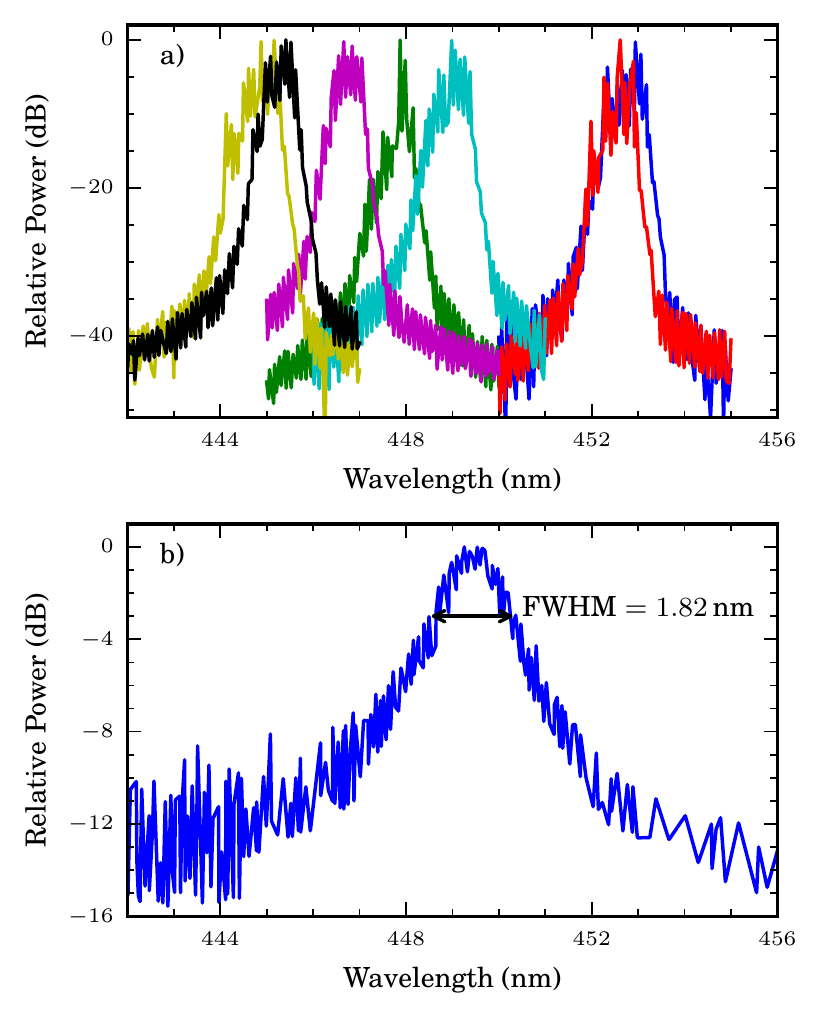}
	\caption{\textbf{a} Normalized spectra of different free-running laser diodes at \SI{60}{\mA}. For a better overview, only 7 from the 12 measured diode spectra are shown. \textbf{b} Spectrum of a laser diode below laser threshold. The FWHM of \SI{1.82}{\nm} is significantly smaller than the one of infrared laser diodes}
	\label{fig:laserdiodes_filter}
\end{figure}

According to the datasheet \cite{OSRAMPl450B}, the center wavelengths of the PL-450B laser diodes spread over a range of \SI{20}{\nm}.
We measured the emission spectra of twelve diodes from different lots and retailers with an optical spectrum analyzer (OSA). It has a wavelength accuracy of \SI{0.05}{\nm} and a resolution of \SI{0.01}{\nm}.
The lasing spectra of a representative subset of seven diodes are shown in Fig. \ref{fig:laserdiodes_filter} a. All spectra were measured with \SI{60}{\mA} current without an external resonator. The central lasing wavelengths cover a range of about \SI{8}{\nm} between \SI{445.15}{\nm} and \SI{452.95}{\nm} with a mean value of \SI{447.66}{\nm}. The measurement also revealed an internal mode separation of \SI{0.05}{\nm} corresponding to \SI{80}{\giga\hertz}.
Figure \ref{fig:laserdiodes_filter} b shows the spectrum of a diode below the laser threshold. The spontaneous emission profile has a FWHM of \SI{1.82}{\nm}, which is much smaller than the \SI{20}{\nm} width of typical infrared diodes.
In addition, the wavelength shift with temperature of \SI{0.06}{\nm/\K} for InGaN \cite{NakamuraChichibu2000} is about five times smaller than for InGaAs.
This requires the preselection of a diode for the targeted wavelength.

\begin{figure}
    \centering
		\includegraphics[]{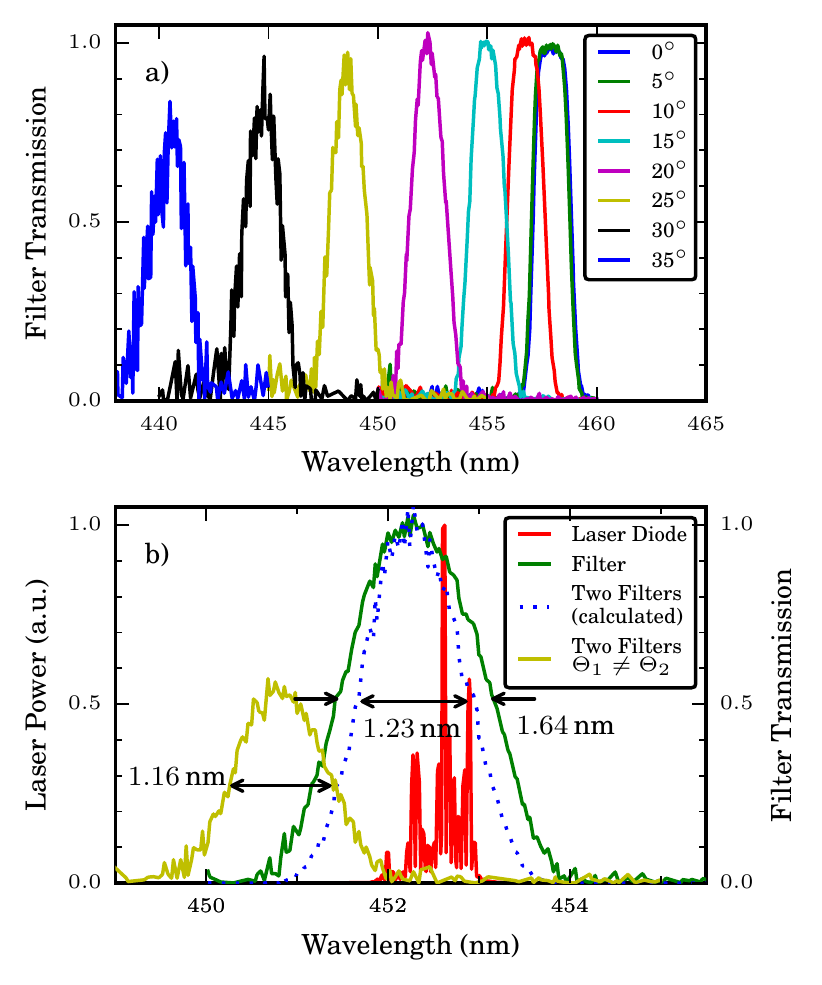}
    \caption{\textbf{a} Spectral transmission of the interference filter for incidence angles between \ang{0} and \ang{35} in steps of \ang{5}. The peak wavelength decreases with increasing angle, and the filter profile changes from a flattop to a Gaussian profile. \textbf{b} Comparison of the emission spectrum of a free-running laser diode (\textit{red}) with filter transmission curves: (\textit{green}) measured transmission of one interference filter at \ang{20}, (\textit{dotted blue}) calculated transmission of two parallel filters at \ang{20} in series, and (\textit{yellow}) two filters in series at $\approx \ang{23}$ with a slight relative tilt ($\theta_1 \neq \theta_2$)}
		\label{fig:laserdiodes_special_filter}
\end{figure}

The spectral transmission of the interference filter (Semrock Laser-Line LL01-458)
was also measured with the OSA using a blue LED as broadband light source. Figure \ref{fig:laserdiodes_special_filter} a depicts the spectral filter transmission for different angles of incidence. As expected, the transmission peak shifts to shorter wavelengths with increasing angle. The maximum transmittance decreases from above \SI{95}{\percent} at \ang{0} to below \SI{80}{\percent} at \ang{35}. At normal incidence the filter has a flattop profile, but at incidence angles of $\theta \ge\SI{20}{\degree}$ the profile is nearly Gaussian.

The FWHM of the transmission stays below \SI{1.8}{\nm} for incident angles $\theta < \ang{25}$.
As shown in figure \ref{fig:laserdiodes_special_filter} b this width is larger than the full emission spectrum of the laser diode and corresponds to the combined spectral width of more than 30 internal cavity modes. These are significantly more modes than in the ECDL setup of Ref. \cite{Thompson2012} in the infrared. Their setup uses an interference filter and laser diode combination, in which only five internal cavity modes fit into the transmission band of the filter. In that case, utilizing the edges of the gain profile of the diode allowed for a stabilization of the laser wavelength. The smaller mode spacing and the fact that the edges of the filters are not as steep as in the infrared contribute significantly to the difficulty of achieving stable single-frequency operation of blue diode lasers. At our targeted wavelength of operation of \SI{452,756}{\nm}, the filter ($\theta \approx \ang{20}$) has a FWHM = \SI{1.64}{\nm}, an edge steepness of \SI{1.01}{\nm} (\SIrange{10}{90}{\percent} value), and a peak transmittance of $\approx$ \SI{98}{\percent}.
To improve the filter performance, we mounted two filter in series. This decreased the FWHM to \SI{1.23}{\nm} and improved the edge steepness to \SI{0.80}{\nm} without a significant loss in transmittance. 
Under small angles of incidence ($\theta < \ang{20}$), a slight variation of the relative angle between the two filters ($\theta_1 \neq \theta_2$), changes the flattop profile of the filter combination to a more Gaussian shape. This reduces the flattop contribution and decreases the FWHM. At higher incident angles ($\theta \ge \ang{20}$), the transmission characteristics is already Gaussian. In this case, a nonzero relative angle between the filters leads to a smaller total transmittance without further improvement in comparison with two parallel filters, as shown in Fig \ref{fig:laserdiodes_special_filter} b. For this reason, we have chosen to align the two filters in parallel.

\subsection{ECDL characterization}
For the determining of the optimum feedback of the external resonator of the ECDL, the intra-cavity beam splitter (BS) was replaced by a polarization beam splitter (PBS) resulting in an effective cavity mirror with adjustable reflectivity and constant polarization: By rotating QWP1, the overall reflectivity of the combination of PBS, QWP1, and mirror M could be changed.
Above \SI{45}{\percent} reflectivity, the laser showed increasing instability. The best stability could be achieved between \SI{10}{\percent} and \SI{20}{\percent} reflectivity, depending on the targeted internal laser mode. A maximum output power of \SI{17}{\mW} could be reached at a wavelength of \SI{454}{\nm}.
Since this configuration does not allow for the implementation of the intra-cavity polarization lock, all further measurements were performed with the ECDL in the original setup described above with beam splitter BS.  
To receive a complete picture, we also tested the ECDL setup from Baillard et al.\ \cite{Baillard2006609} with one filter and a \SI{15}{\percent} back reflecting mirror used for feedback and outcoupling. Here, an output power of up to \SI{22}{\mW} was achieved, but single-frequency operation was limited to a duration of \SI{1}{\hour} typically.

\begin{figure}
	\centering
	\includegraphics[width=\linewidth]{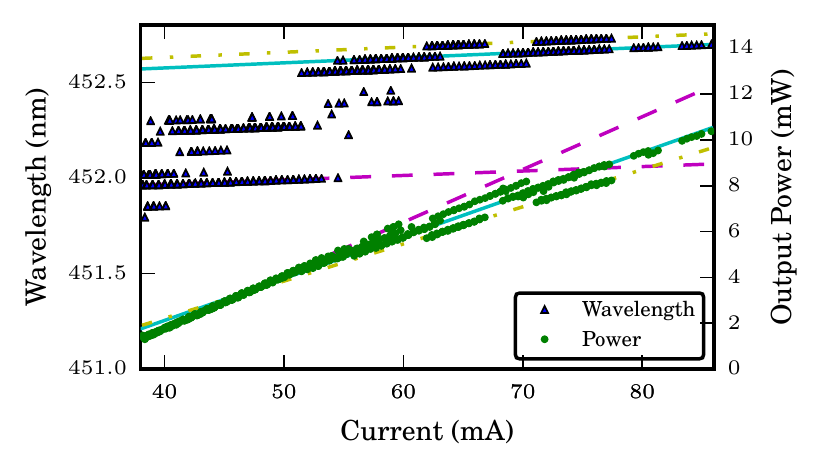}
	\caption{Output power and wavelength characteristics of the two-filter ECDL. The wavelength jumps in discrete steps of \SI{0.05}{\nm} between the internal cavity modes of the laser diode. A specific mode shows a shift of \SI{2.59 +- 0.04}{\pico\m/\mA}. The slope of the corresponding output power is dependent on the specific internal mode. This correspondence is indicated by three pairs of linear fits with matching line style and color (\textit{solid cyan}, \textit{dashed magenta}, \textit{dash-dotted yellow}) for power and wavelength}
	\label{fig:laser_characteristics}
\end{figure} 

To further characterize the setup, the wavelength and the output power of the double-filter ECDL ($\theta \approx \ang{20}$) were measured as a function of the diode injection current (see figure \ref{fig:laser_characteristics}).
The wavelength was measured with a commercial wavemeter and the optical output power with a calibrated photodiode. 
With increasing current, the wavelength and the output power increase. 
The laser diode current was changed in steps of \SI{21}{\uA}. Data were only recorded when the wavemeter confirmed single-mode operation of the laser. Below \SI{38}{\mA} and above \SI{86}{\mA} no stable operation was achieved. 
Following the internal modes (indicated by linear fits with specific colors), the output power rises linearly with the applied current, with different slopes for different modes.
More than \SI{10}{\mW} of output power could be achieved at \SI{86}{\mA} injection current.
The spatial profile of the output mode resembles an elliptical Gaussian TEM$_{00}$ mode with a 2.5:1 ratio of the spatial widths. We achieve a coupling efficiency of about \SI{60}{\percent} of the output mode into a single-mode optical fiber.
The slope of the wavelength shift for all laser modes is similar at about \SI{2.59 +- 0.04}{\pico\meter / \mA}.

With rising current, additional jumps in the wavelength of \SI{0.05}{\nm} towards longer wavelengths occur. The step size agrees with the internal mode spacing of the laser diode.
Due to the broad filter transmission compared to the internal mode spacing, the wavelength can be tuned over a wide range of \SI{0.8}{\nm} without changing the angle $\theta$ of the filters. The stability is highest at high currents: Here, the laser medium favors emission at longer wavelengths, whereas the edge of the filters suppresses internal modes with too large wavelengths.
The measurement also shows a current range of several tens of \si{\uA} for stable operation at a given laser mode before the next mode hop occurs. Because of this small range, we use a digital current driver which allows us to vary the current in steps of \SI{2}{\uA}.

\subsection{Frequency Stability}
\begin{figure}[tb]
	\centering
	\includegraphics{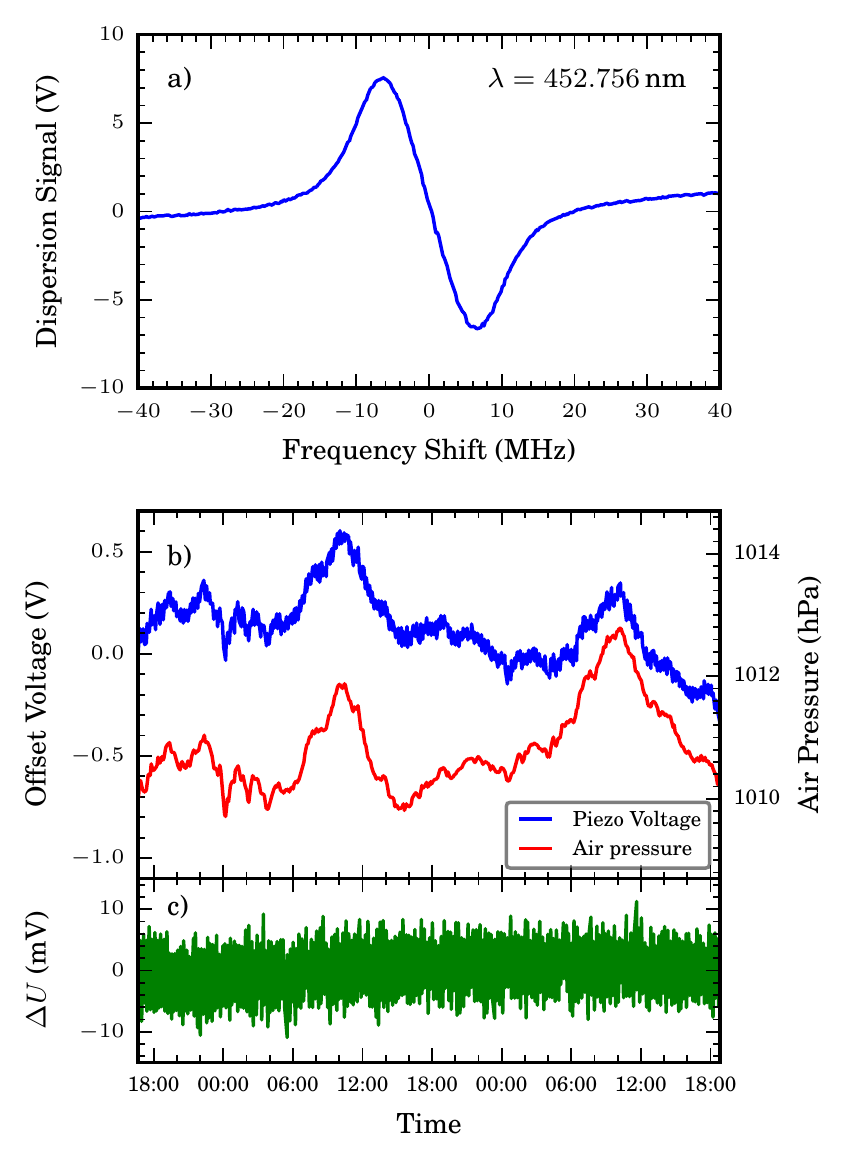}
	\caption{\textbf{a} Measured Doppler-free dispersion spectrum of an absorption line of molecular tellurium at \SI{452.756}{\nm}. \textbf{b} Long-term stability of the laser recorded over a period of two days. The laser is stabilized to the Doppler-free tellurium line shown in \textbf{a}. The applied offset voltage of the ring piezo compensates the length change of the external resonator predominantly caused by air pressure variations. \textbf{c} The corresponding error signal of the tellurium stabilization loop demonstrates the stable laser operation over the full measurement period}
	\label{fig:longtermstability}
\end{figure}

Due to the broad transmission spectrum of the utilized interference filters compared to the laser diode mode spacing, the frequency stability of the ECDL is strongly dependent on the external cavity configuration. With the original setup of Ref. \cite{Baillard2006609} and one filter, single-frequency operation could be sustained for up to \SI{1}{\hour}. Switching to the two-filter configuration described in section \ref{sec:setup} without intra-cavity locking applied, extended the duration of single-frequency operation to several hours. The laser remained sensitive to small disturbances. These can lead to a misalignment between internal and external resonator, which causes mode jumps. 
A significant improvement of the long-term stability could be achieved by adding the intra-cavity polarization locking scheme described above. 
For a characterization in the long-term stability, the laser was frequency-stabilized via Doppler-free saturation spectroscopy to $^{130}$Te$_{2}$ which has a dense line spectrum in the blue wavelength region \cite{cariou1980atlas}. 
The stabilization on a Lamb-dip of tellurium uses a second lock-in servo loop, which generates the derivative of the Lamb-dip via a \SI{40}{\kHz} modulation of the laser current. This modulation is faster than the feedback bandwidths of both lock-in servo loops.
Figure \ref{fig:longtermstability} a shows a scan of the dispersion signal of the targeted tellurium Lamb-Dip at \SI{452.756}{\nm} which is an unlabeled line about halfway between the lines labeled 521 (\SI{22085.9292}{\per\cm}) and 522 (\SI{22087.6258}{\per\cm}) in \cite{cariou1980atlas}. With a second PI controller, this signal is fed to the piezo of the ECDL cavity and to the laser diode current. 

For a long-term measurement of the frequency stability, the tellurium lock error signal and the piezo voltage for stabilizing the ECDL cavity length were recorded in \SI{5}{\s} intervals. For each interval, the signals were averaged over 100 samples to reduce the noise. In order to determine the influence of the laboratory conditions, room temperature, ambient air pressure, and humidity were recorded every five minutes.
Figure \ref{fig:longtermstability} b,c shows a respective measurement over a duration of two days.
The laser stays locked over the full duration and thus demonstrates excellent long-term stability. 
The piezo voltage shows a strong correlation with air pressure, superimposed by a small negative piezo voltage drift, starting after the first half of the measurement time.
The room temperature of \SI{21.5}{\celsius} was stable to better than \SI{0.5}{\K} and the relative humidity stayed between \SI{40}{\percent} and \SI{50}{\percent}.
The absolute variation of the piezo voltage does not exceed \SI{1}{\V} which corresponds to an upper bound for the change of the cavity length of \SI{100}{\nm}.
Figure \ref{fig:longtermstability} c confirms that the laser stayed in lock during the full measurement time. The in-loop deviation of the error signal is symmetric around the setpoint of \SI{0}{\V} with a mean excursion of only a few mV. An out-of-lock event would show a large discontinuous jump of several tens of \si{\mV}.

\section{Conclusion}
We built and characterized an external cavity diode laser setup with two inexpensive, commercially available interference filters giving an optical output power of more than \SI{10}{\mW} at about \SI{450}{\nm}.
The use of two filters instead of one for wavelength selection in the external resonator increased the edge steepness and thus enhanced the stability of single-frequency operation significantly. In addition, the configuration of the external resonator with a beam splitter cube, a quarter wave plate, and a highly reflective mirror allowed for fine-adjustment of the back-reflectivity and for the implementation of an intra-cavity locking scheme for matching the diode laser modes to the ECDL modes. 
This scheme increased the duration of stable single-frequency operation from a few hours to more than two days. 
This could be demonstrated by stabilizing the laser to a tellurium absorption line at \SI{452.756}{\nm}.

To further optimize the stability of the ECDL, the length of the external resonator can be decreased, reducing the effects of air pressure and temperature variations.
If a higher output power is needed, the measured optimal feedback condition can be implemented by building an ECDL with two filters and a reflectivity optimized semitransparent output mirror in cateye configuration with less intra-cavity losses. The intra-cavity locking scheme still can be applied. For a reflectivity between \SI{10}{\percent} and \SI{20}{\percent}, an output power above \SI{20}{\mW} is achievable. 

\section*{Acknowledgments} 
We dedicate this contribution to Ted Hänsch in honor of his 75th birthday. Without his seminal contributions to physics in general, laser science and technology, atomic physics, and quantum optics in specific, modern physics would not be as fascinating as it is, and the work reported here would not have been possible. 
We wish to thank Holger John and the group of Thomas Walther for fruitful discussions and advise on the laser design and the support with the OSA measurements. We appreciate the contributions of Florian Vollrath to the stabilization system. 
This work received partial financial support from the Deutsche Forschungsgemeinschaft (DFG) under the Grant No.\ BI 647/4-1. A.\ Martin and P. Baus acknowledge support from HGS-HIRe. We thank the ARTEMIS collaboration for continuing support. The experiments have been performed within the framework of the HITRAP facility at the Helmholtz Center for Heavy Ion Research (GSI) and the Facility for Antiproton and Ion Research (FAIR) at Darmstadt.

\bibliographystyle{osajnl_url}
\bibliography{450nmLaserLit}

\end{document}